\documentclass{elsart}
\journal{}

\usepackage{epsfig}
\usepackage{amssymb}
\usepackage{amssymb}
\usepackage{amsmath}
\usepackage{textcomp}

\begin{document}


\begin{frontmatter}

\title{Pulse shape simulation for segmented true-coaxial HPGe detectors}

\author[a]{I.~Abt\corauthref{cor}}\ead{isa@mppmu.mpg.de},
\author[a]{A.~Caldwell}, 
\author[a]{D.~Lenz},
\author[a,b]{J.~Liu}, 
\author[a]{X.~Liu},
\author[a]{B.~Majorovits}
\address[a]{Max-Planck-Institut f\"ur Physik, M\"unchen, Germany}
\address[b]{now: Institute for the Physics and Mathematics of the Universe, Tokyo University}
\corauth[cor]{Max-Planck-Institut f\"ur Physik, F\"ohringer Ring 6, 
              80805 M\"unchen, Germany, 
              Tel.: +49-(0)89-32354-262, FAX: +49-(0)89-32354-528}

\begin{abstract}

A new package to simulate the formation of electrical pulses
in segmented true-coaxial high purity germanium detectors is
presented. The computation of the electric field and 
weighting potentials inside the
detector as well as of the trajectories of the charge carriers
is described.
In addition, the treatment of bandwidth limitations and noise
are discussed.
Comparison of simulated to measured pulses, obtained
from an 18-fold segmented detector operated inside a cryogenic
test facility, are presented.

\end{abstract}

\end{frontmatter}

\section{Introduction}
\label{section:introduction}

High purity germanium detectors, HPGeDs, are used in a wide variety 
of applications in particle and nuclear physics\,\cite{Kno99,eberth}. 
The analysis of their pulse shapes, PSA, can be a powerful tool 
to determine event topologies,
often a key to distinguish between
signal and background events\,\cite{psam,photon,psa,scs}. 
In particular, the separation of events with energy depositions in
one or multiple locations inside an HPGeD, single-site or multiple-site
events, is of interest.
Previous studies were often limited by the availability of clean 
event samples to train neural nets or to evaluate PSA\,\cite{psa,scs}. 
Therefore, simulated events reflecting a detailed understanding of
the underlying processes are extremely valuable.

The interactions of radiation ($\alpha, \beta, \gamma, n, p$, etc.)  
inside a semiconductor  create electron-hole pairs.
These charge carriers are separated by the electric field 
inside an HPGeD and drift towards the electrodes
inducing time dependent charges.
In most applications, charge sensitive amplifiers are used and
the resulting pulses are sampled and digitized
at a given frequency.

A new pulse shape simulation, PSS, package was developed.
The geometry of an 18-fold,$(6\phi,3z)$, segmented true-coaxial
HPGeD~\cite{si} is implemented as default.
The inner mantle has a radius of 5\,mm, the outer of 37.5\,mm.
The length is 70\,mm. 
This allows the comparison to data obtained from
such a detector which was developed in connection to the
GERDA~\cite{gerda} experiment.
However, the simulation package itself is general and 
the geometry can easily be modified to accommodate
any kind of germanium detector.

\section{Simulation procedure}
\label{s:proc} 

The package\,\cite{jing,daniel} is separated into two parts, the
calculation of the static properties of the
HPGeD followed by the event by event simulation of the pulse development.
The tabulation of the static properties is done once at the beginning
of the simulation. It includes the electric field and the weighting
potentials inside the detector. 

The event by event simulation comprises the event topology
and the development of the pulses in time-steps:

\begin{enumerate} 
\item Simulation of the interactions of particles with germanium.
      The output are individual positions with energy depositions, hits; 

\item Clustering of hits.
      According to the requirements of the simulation, hits can be
      clustered~\footnote{Hits closer to each other than the best
                hypothetical radial resolution of the detector, 1\,mm,
                calculated as the speed of the charge carriers multiplied 
                by the time resolution of the electronics,
                are clustered by default.}.
      The position of the cluster is the 
      bary-center of the original hits. The energy of the 
      cluster, $E_{\mbox{cl}}$, is the sum of the energies of the original 
      hits. 

\item Simulation of the drift of charge carriers in time-steps. 
      For each cluster, only one point-like charge is considered.
      At each time-step the velocity of the charge is calculated
      using the electric field as tabulated;

\item Calculation of the charges induced in the electrodes after each step;

\item Simulation of experimental effects such as noise, bandwidth
      limitation, shaping times, etc..  
\end{enumerate}
 
The first step is done using the {\sc GEANT4}\,\cite{G403,G406} based 
simulation package {\sc MaGe}\,\cite{MaGe}, jointly developed by the GERDA 
and Majorana collaborations. The good agreement between MaGe 
and data obtained with segmented detectors was demonstrated
previously~\cite{photon}.

All calculations are performed in cylindrical coordinates,$r, \phi$ and $z$, 
with the origin at the center of the detector.
 
\section{Electric field and weighting potentials} 
\label{s:field} 

The electric field, $\vec{E}(x)$, at a position~$x=(r,\phi,z)$ 
inside an HPGeD
depends on the geometry of the detector, 
the bias voltage, $V_{bias}$ applied 
and the density of electrically active impurities,
$\rho_{imp}(x)$.
It is considered static, calculated only once and tabulated
for a reasonably spaced grid~\footnote{The default is a grid with
a 1\,mm spacing in $r$ and $z$ and a 2\,degree spacing in~$\phi$.}.
The electric field at any position 
is calculated by interpolation.

For simple cases like a
constant $\rho_{imp}(x)$ , 
$\vec{E}(x)$ can be calculated analytically solving
Poisson's equation 
  $\nabla \cdot \vec{E(x)} = \frac{\rho(x)}{\epsilon}$, 
where $\rho(x)$ is the space charge density 
$\epsilon$ is the dielectric constant. 
However, the electric field 
is always calculated numerically using 
the potential $\varphi(x)$ and the boundary conditions. 

The boundary conditions for $\varphi$ depend on $V_{bias}$.
For the true-coaxial $n$-type HPGeDs considered here, 
the potential is fixed to $ V_{bias}$ on
the inner mantle, zero on the outer mantle and it floats on the end 
surfaces. 
The density $\rho_{imp}(x)$ determines $\rho(x)$ 
and thereby affects 
the electric field inside an HPGeD.
In the case of constant $\rho_{imp} $, the 
electric field only depends on $r$.
For this case, the output of the numerical calculation based on successive 
over-relaxation was tested against the numerical solution~\cite{daniel}.
The deviations for realistic values of $\rho_{imp}(x)$ 
were found to be less than 0.6\,\% throughout the detector. 

\begin{figure}[htbp] 
\centering 
\includegraphics[width=0.9\linewidth]{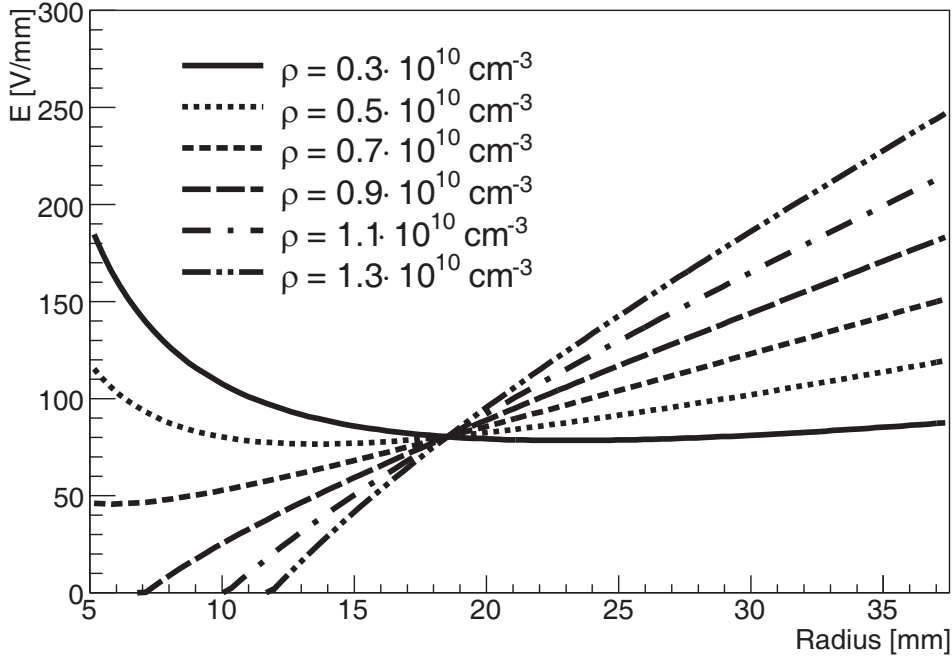} 
\caption{Strength of the electric field as a function of the 
cylindrical coordinate $r$ for constant impurity densities 
between 0.3 and $1.3 \times 10^{-10}$/cm$^{3}$.} 
\label{f:rho} 
\end{figure}

Figure~\ref{f:rho} shows the 
calculated strength of the electric field  
as a function of $r$ for $V_{bias} =$~3\,kV and varying 
constant $\rho_{imp}$.
This range of $\rho_{imp}$ values is realistic for HPGeDs.
For lower $\rho_{imp}$, the  field is strong throughout the
detector.
For higher $\rho_{imp}$, the field 
is insufficient at small radii; a higher  $V_{bias}$ is needed.

As $\rho_{imp}(x)$ can vary up to a factor three between the two ends 
of a detector and might depend on $r$, it is possible to specify
a three dimensional distribution for $\rho_{imp}(x)$.

The influence of any electrode 
on a given space-point inside the HPGeD is characterized by  
its weighting potential.
This is defined according to
Shockley-Ramo's theorem~\cite{Gat82,Rad88,He00}
as the solution of Poisson's equation
for the boundary conditions that the 
potential on the electrode of interest equals unity and the potentials 
on all other electrodes equal zero.  
All weighting potentials are treated numerically like the electric field.

\section{Drift of charge carriers} 
\label{s:drift} 

The drift of the charge carriers is calculated in time-steps, $\Delta t$,
for each point representing a cluster of hits as defined in 
Section~\ref{s:proc}.
At each step the velocity is calculated using the electric field as tabulated.
Between grid points a simple linear extrapolation is used for the electric
field.
Two different numerical methods are implemented to calculate the 
trajectories, the Euler method and the 4$^{th}$~Runge-Kutta method. 
The former is less computer time intensive, but is also less precise. 
However, for step-sizes $\Delta t \lesssim 1$~ns, the resulting trajectories
do not differ significantly\,\cite{jing}.  


The drift velocity of the charge carriers, 
$\vec{v}_{e}(x)$ for electrons and $\vec{v}_{h}(x)$ for 
holes, is calculated for each step 
using the electric field, $\vec{E}(x)$,
as tabulated:
\begin{equation} 
\label{e:dv} 
\vec{v}_{e/h}(x)= \mu_{e/h}(x) \vec{E}(x), 
\end{equation} 
where $\mu_{e/h}(x)$ is the mobility~\cite{Kno99}  
of electrons and holes, respectively.  
 
Whether $\mu_{e/h}(x)$ depend on the relative position of $x$
to the crystal axes is determined
by the temperatures of the germanium crystal, $T_{crystal}$,
and of the charge carriers, $T_{cc}$.  
If $T_{crystal} \approx T_{cc}$, $\mu_{e/h}(x)$ 
become numbers, $\mu_{e/h}^0$
and the drift velocities are
directly proportional to the electric field. 
However, HPGeDs are normally operated at
temperatures around 100\,K and  $T_{crystal} <<  T_{cc}$.
In this case,the mobility becomes a complex tensor.
Therefore, the simulation has to take the crystal structure into account
and the calculation of $\vec{v}_{e/h}(x)$ becomes quite involved.
The drift of the charge carriers is not any longer parallel to
the electric field~\cite{mobilitytensor} for all  $x$.

The axes of the face-centered cubic structure
of germanium crystals are characterized    
by Miller indices~\cite{millerindices}.
Cylindrical germanium detectors are produced with their 
$z$ axis aligned to the crystal axis $\langle 001 
\rangle$~\cite{z-align}. This is implemented in the simulation.
The axes $\langle 100 \rangle$ and $\langle 110 \rangle$ are in the
$r\phi$-plane. Their relative position to the segment boundaries
can be chosen freely, i.e. adjusted to the detector
to be simulated.

The tensors $\mu_{e/h}(x)$ are unfortunately not known.
Special measurements along the crystallographic axes are used and
parameterized. The results are combined to get results for
any point in the crystal.

If the electric field is parallel to any of the principal 
crystallographic axes, the charge carriers will drift along this axis, 
because of the symmetric structure of the 
germanium crystal~\cite{mobilitytensor,axissymm}. 
The measured drift velocities~\cite{bart,miha,reg}, $v_{e/h}$,
along the axes $\langle 100 \rangle$ and $\langle 111 \rangle$ with 
$\vec{E}(x)\parallel \langle 100 \rangle$ for $x$ on 
$\langle 100 \rangle$
and
$\vec{E}(x)\parallel \langle 111 \rangle$ 
for $x$ on 
$\langle 111 \rangle$,
respectively,
are parametrized well~\cite{Kno99,miha} 
for $E=|\vec{E}(x)| < 300$\,V/mm by

\begin{equation} 
\label{e:para} 
v_{e/h} = \frac{\mu_{e/h}^{0}E(x)}
           { [1+(\frac{E(x)} {E_0} )^{\beta}]^{1/\beta} }, 
\end{equation}

where $\mu_{e/h}^{0}$, $E_0$ 
and $\beta$ are parameters determined by fitting. 
The parameters $\mu_{e/h}^{0}$
represent the linear relation between $\vec{v}$ and $\vec{E}$ 
at large $T_{crystal}$ and low $E$. 
The parameters 
$E_0$ and $\beta$ are used to
model the deviation from this linear 
relation at low lattice temperature and high electric fields.  
 
The set of parameters given in~\cite{bart} and  
validated for the drift of electrons~\cite{bart2} 
was used in the simulation presented here. The values are 
listed in Table~\ref{t:pars}.

\begin{table}[htpb] 
\centering 
\caption{Parameters for the experimental drift velocities in the 
$\langle111\rangle$ and $\langle 100 \rangle$ directions 
used in the simulation.} 
\label{t:pars} 
\begin{tabular}{ccccc} 
\hline\noalign{\smallskip} 
Carrier & Axis & 
$\mu_{0} \left[\frac{\mbox{cm}^{2}}{\mbox{V}\cdot\mbox{s}}\right]$ & 
$\mathcal{E}_{0}\left[\frac{\mbox{V}}{\mbox{mm}}\right]$ & $\beta$\\ 
\noalign{\smallskip}\hline\noalign{\smallskip} 
 
{$e$} & $\langle111\rangle$ & 38536 & 53.8 & 0.641 \\ 
      & $\langle100\rangle$ & 38609 & 51.1 & 0.805 \\ 
\hline 
 
{$h$} & $\langle111\rangle$ & 61215 & 18.2 & 0.662 \\ 
      & $\langle100\rangle$ & 61824 & 18.5 & 0.942 \\ 
\noalign{\smallskip}\hline\noalign{\smallskip} 
 
\end{tabular} 
\end{table} 
 
\begin{figure*}[tb] 
\centering 
\includegraphics[width=0.45\linewidth]{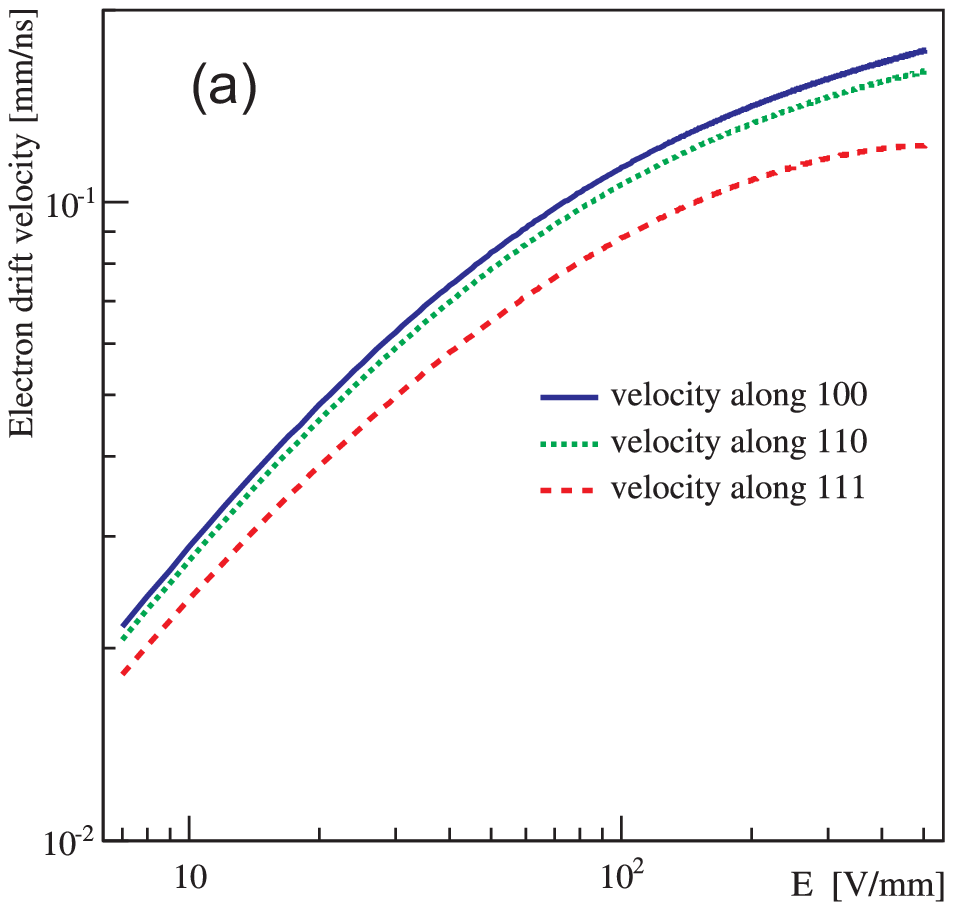} 
\includegraphics[width=0.45\linewidth]{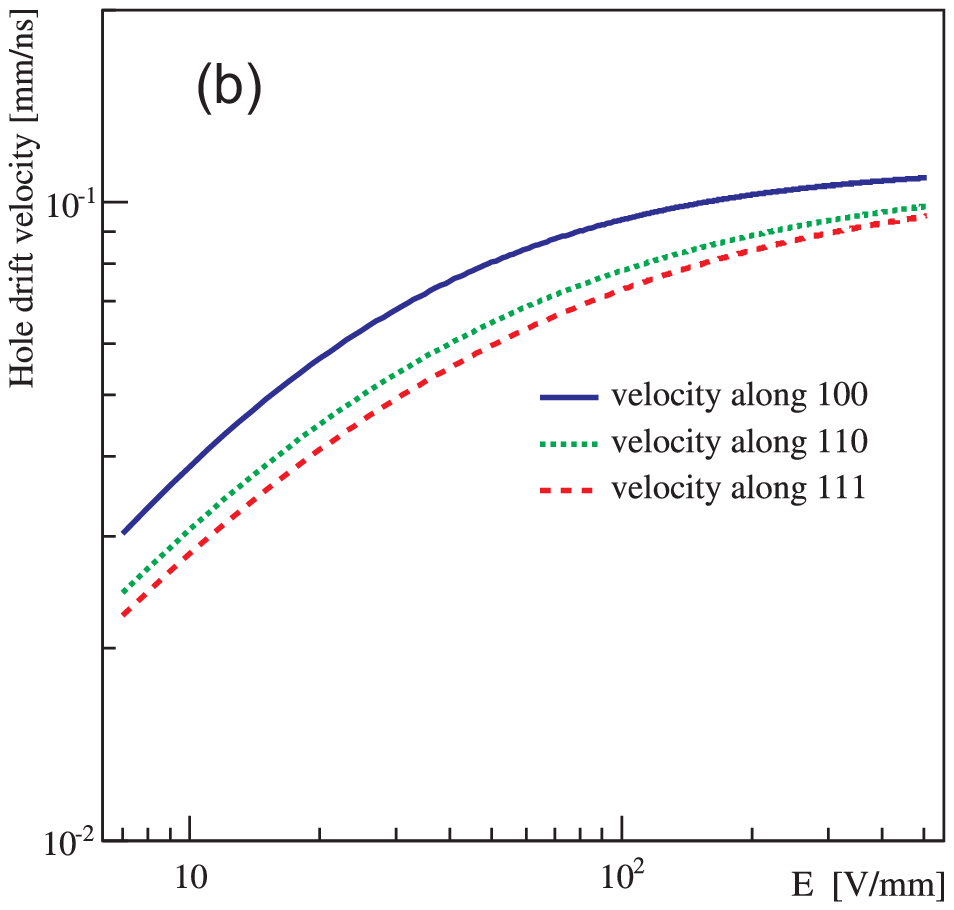} 
\caption{Drift velocities of (a) electrons and (b) holes along 
the principal crystal axes as function of the electric 
field. Velocities along the axes $\langle 111 \rangle$ and $\langle 
100 \rangle$ are calculated according to Eq.~\ref{e:para}. 
For the calculation along  $\langle 111 \rangle$, see text.
} 
\label{f:vvse} 
\end{figure*} 
 

The drift velocities in any direction can be derived from the 
velocities along the $\langle 111 \rangle$ and $\langle 100 \rangle$
axes.  
The model used for the electron drift\,\cite{miha} is based on the 
idea that the conduction band in a germanium crystal reaches its 
minimal potential in regions around the four equivalent $\langle 111 
\rangle$ axes.  
Free electrons effectively only populate these regions.
The probability density of free electrons in other 
regions can be ignored. 
The resulting drift velocities along 
the  $\langle 110 \rangle$ axis are shown together
with the velocities for the
$\langle 111 \rangle$ 
and
$\langle 100 \rangle$  
axes in~Fig.~\ref{f:vvse}(a).

The model used to calculate the hole drift velocities~\cite{bart} in any 
direction
is based on the idea 
that only the ``heavy hole valence band''\,\cite{heavy} is 
responsible for the anisotropy of the mobility. All other effects are 
neglected.   
A parameterization~\cite{bart}
is used to calculate the three components of the hole drift velocity $\vec{v}$
at any position.  
The results for the  $\langle 110 \rangle$ axis
together with the values for the
$\langle 111 \rangle$ 
and
$\langle 100 \rangle$  
axes are shown in~Fig.~\ref{f:vvse}(b).

\section{Trajectories} 
\label{s:trj} 

The results presented 
here were obtained with the Runge-Kutta method. 
\begin{figure*}[tb] 
\centering 
\includegraphics[width=0.9\linewidth]{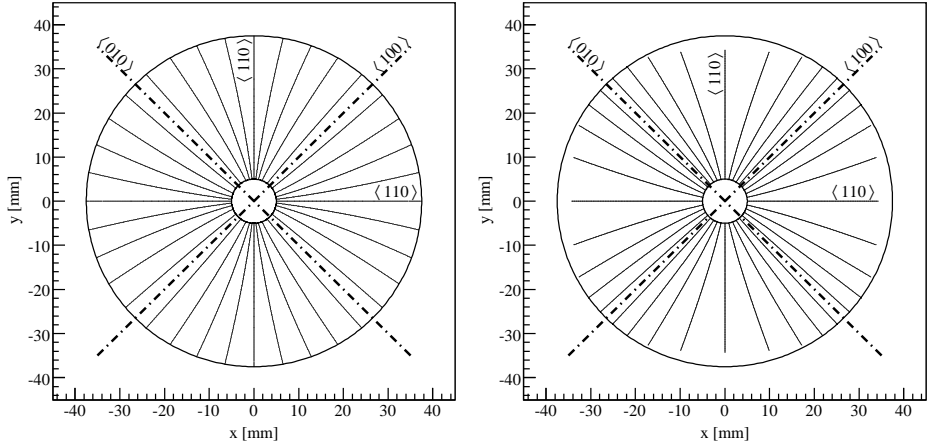} 
\caption{Trajectories of (left) electrons and (right) 
holes in the $(x,y)=(r,\phi)$
plane for constant $\rho_{imp}$.
 Electrons drift inwards and holes outwards.
} 
\label{f:trjs} 
\end{figure*} 
 
Figure~\ref{f:trjs} shows the trajectories of electrons and holes
created at the outer and inner mantle, respectively.
The step size was set to 1\,ns and the total time simulated 
was 400\,ns, $V_{bias}$\,=\,3000\,V and $\rho_{imp} = 0.62 \cdot 10^{10}$
was constant
through the volume.
The trajectories are bent due to the crystal structure;
this is called transverse anisotropy. 
The charge carriers are slower in the $\langle 110 \rangle$ than
in the $\langle 100 \rangle$ direction; this is called longitudinal
anisotropy.
As holes are slower than electrons, the longitudinal anisotropy can be seen
very clearly along $\langle 110 \rangle$ where the holes do not reach
the outer mantle within 400\,ns. 

\begin{figure}[tb] 
\centering 
\includegraphics[width=0.9\linewidth]{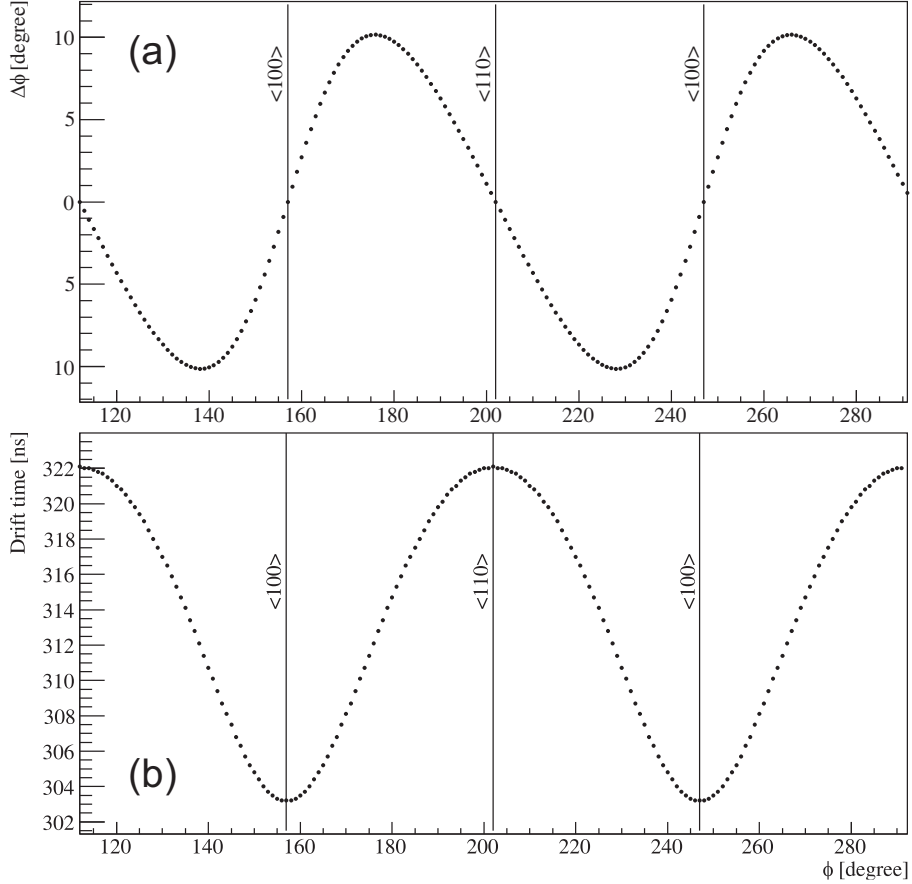} 
\caption{(a) Displacement $\Delta\phi$
between starting point on the outer mantle and the end point 
on the inner mantle of the detector as a function of the
azimuthal angle $\phi$ at the start.(b) Total drift-time as a function of
$\phi$.
}
\label{f:dphi} 
\end{figure} 

The transverse anisotropy is also demonstrated for electrons in 
fig.~\ref{f:dphi}. It shows the displacement $\Delta\phi$
at the end of the drift 
and the total time  needed for the drift 
as functions of the azimuthal angle $\phi$ at the start.
Along the crystal axes $\langle 110 \rangle$ and $\langle 100 \rangle$,
$\Delta\phi = 0$. The drift time is minimal at $\langle 100 \rangle$
and maximal at $\langle 110 \rangle$.

\section{Charges induced  on electrodes} 
\label{s:wei}

\begin{figure*}[tb] 
\centering 
\includegraphics[height=0.35\textheight]{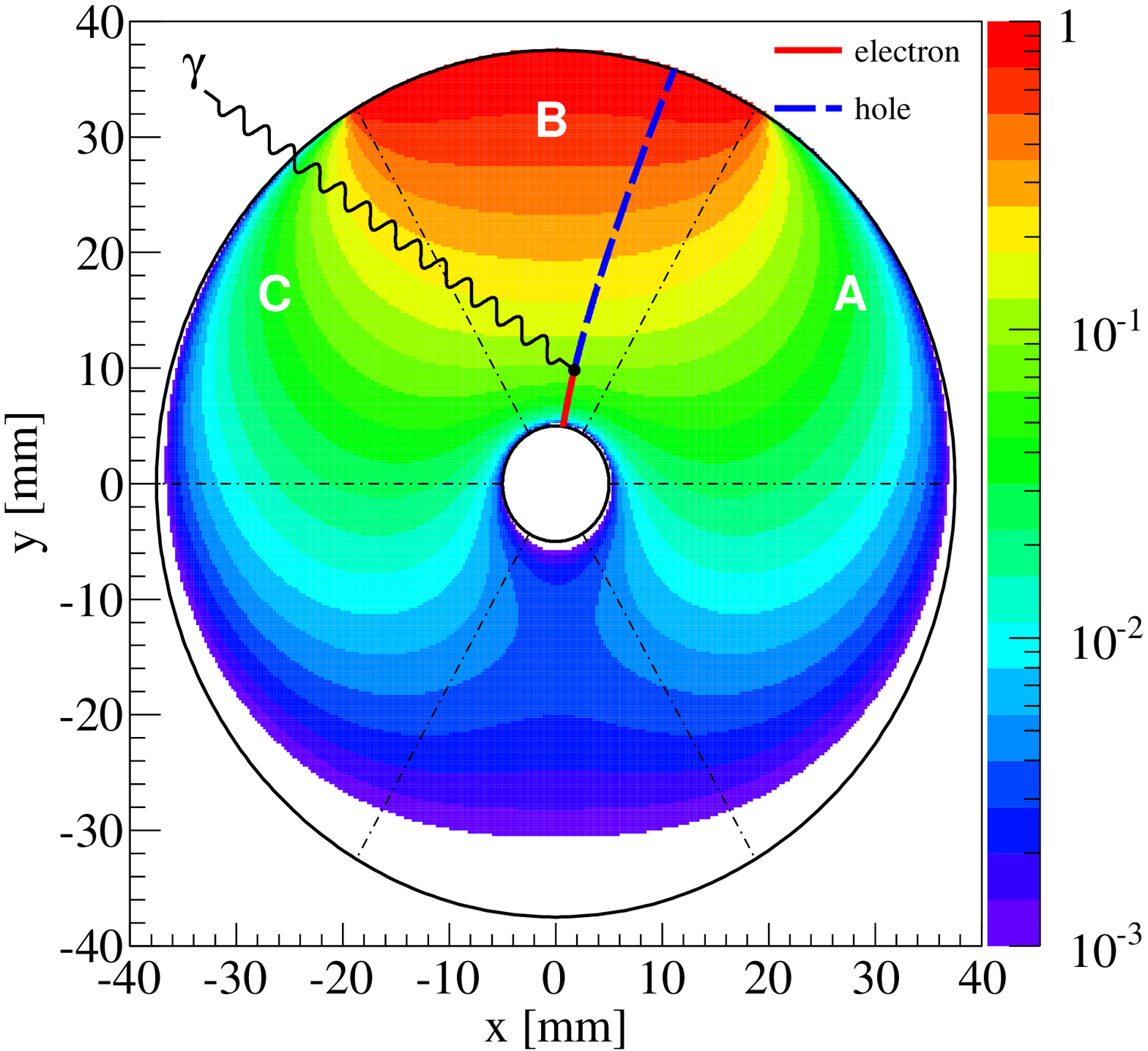} 
\vskip 2mm
\includegraphics[height=0.27\textheight]{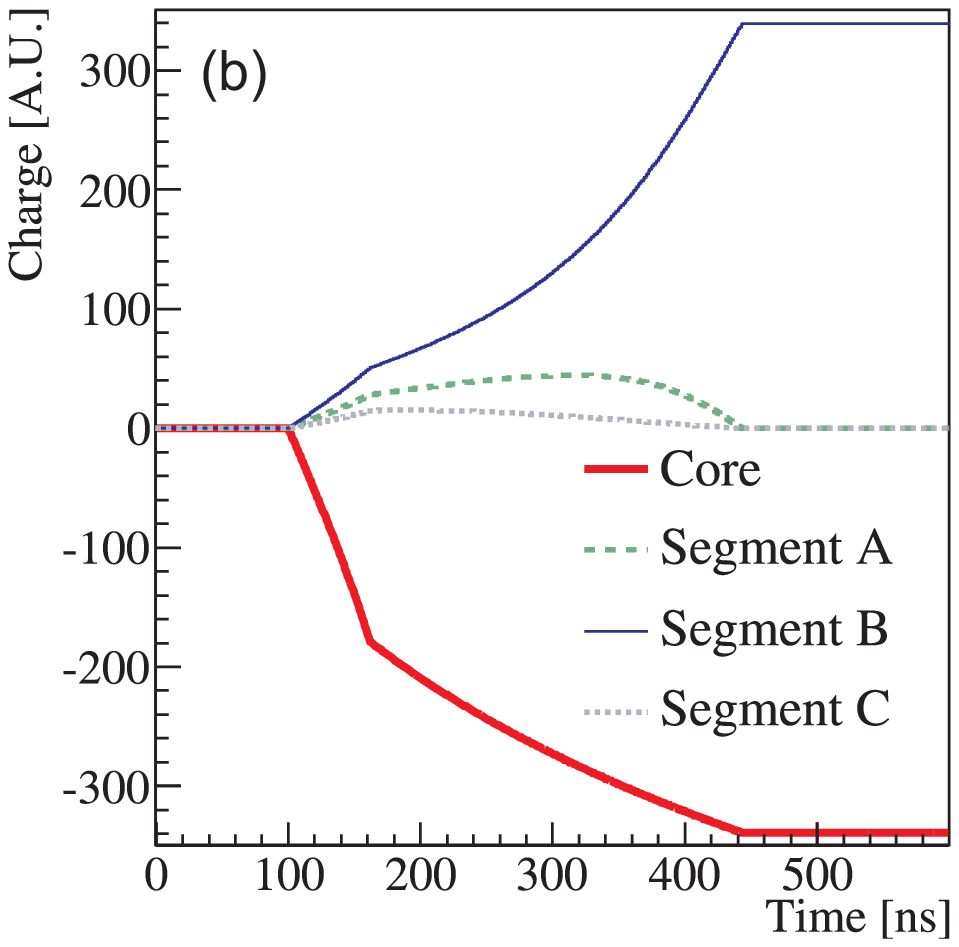} 
\includegraphics[height=0.27\textheight]{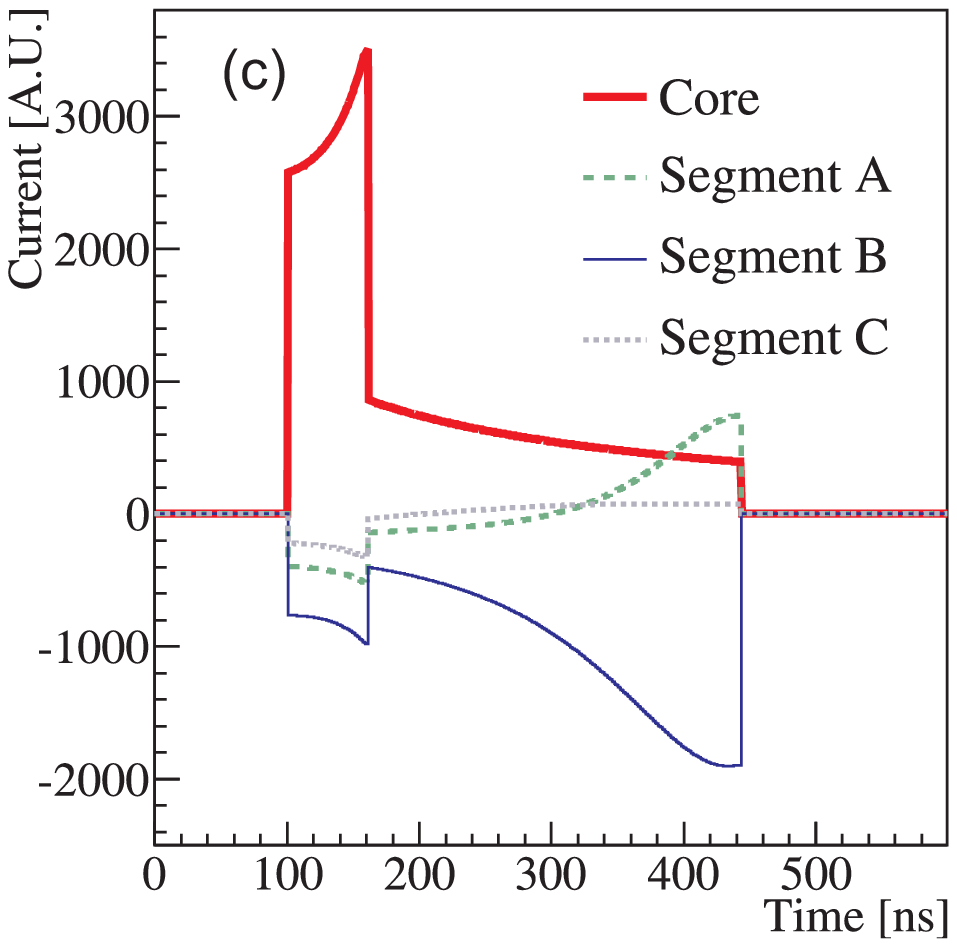} 
\caption{(a) weighting potential of segment B together with an 
indication of a $\gamma$ interaction. (b) Simulated charge and (c) current 
pulses induced in segments A, B and C and the core.} 
\label{f:ps} 
\end{figure*}

\begin{figure*}[tb] 
\centering 
\includegraphics[width=0.85\textwidth]{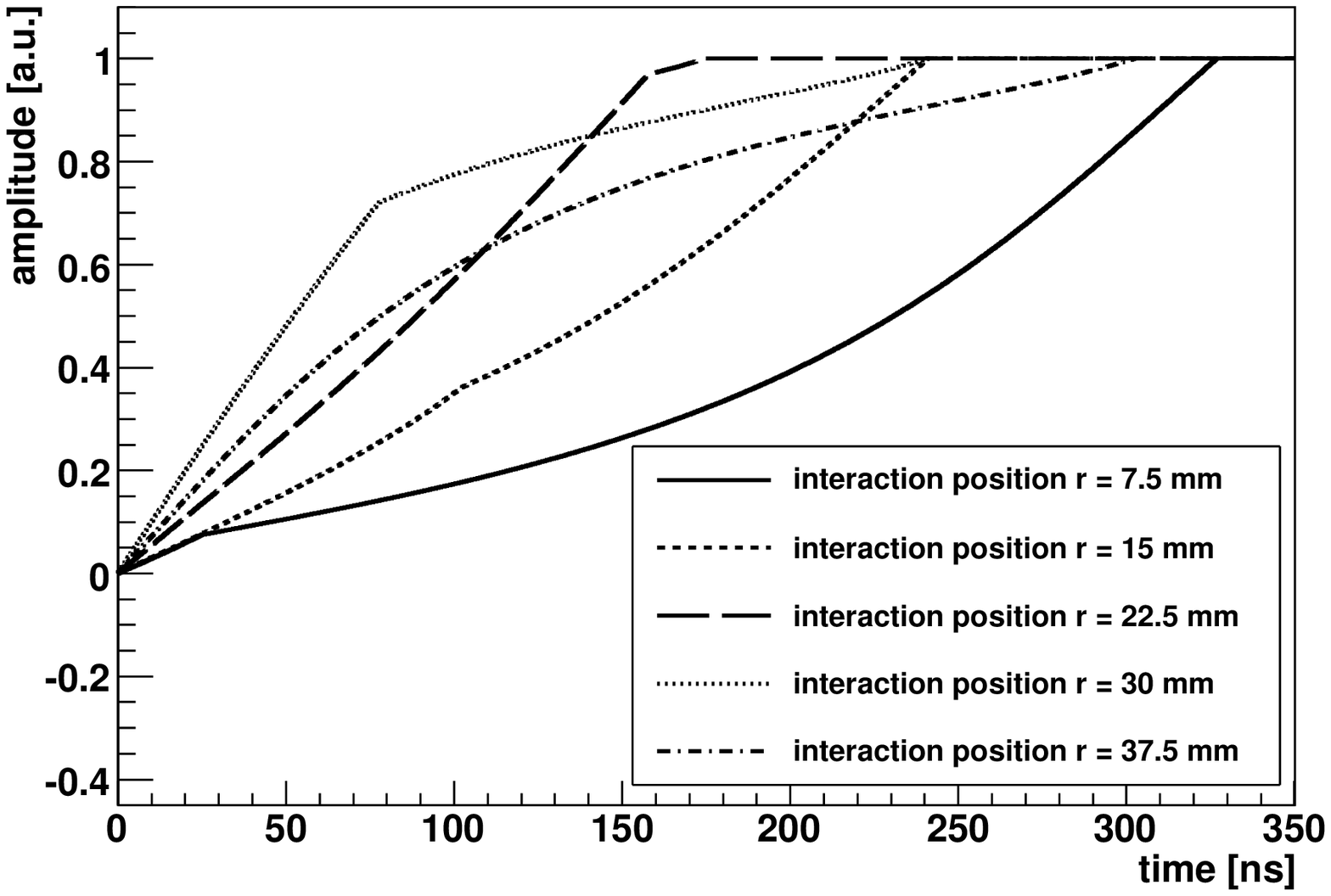} 
\caption{Simulated charge pulses in the segment 
electrode for charges deposited at different radius, $r$.}
\label{f:psrad1} 
\end{figure*} 
\begin{figure*} 
\centering 
\includegraphics[width=0.85\textwidth]{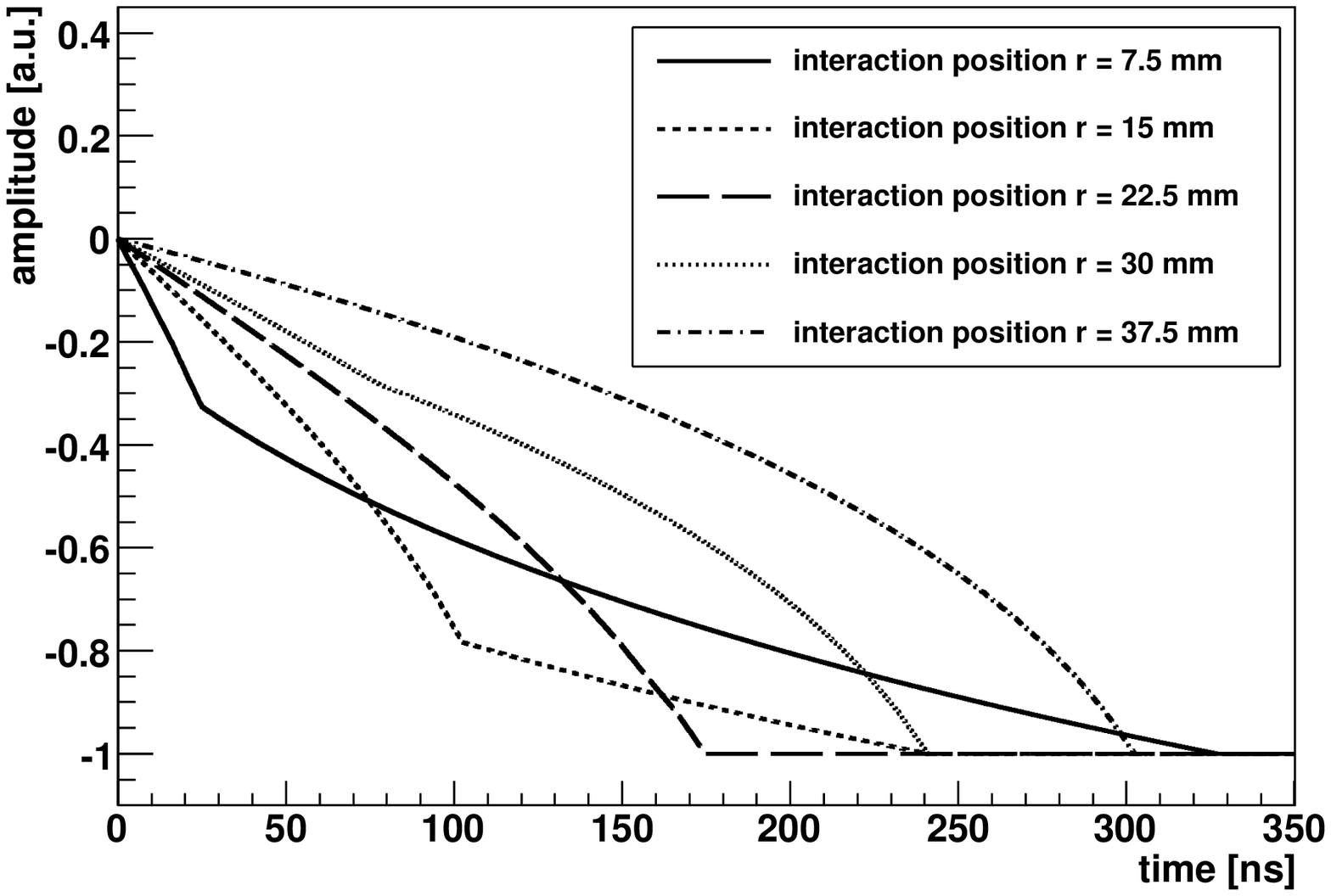} 
\caption{Simulated charge pulses in the
core electrode for charges deposited at different radius, $r$.}
\label{f:psrad2} 
\end{figure*} 

The electric signals are induced in the electrodes of a detector by the 
cumulative influence of moving electrons and holes.
Shockley-Ramo's Theorem \cite{Gat82,Rad88,He00} is 
used to calculate the time development of the induced charge $Q(t)$ or 
current $I(t)$ in each electrode: 
\begin{equation} 
\label{e:ramoq} 
Q(t) = -Q_{0} \cdot [\varphi_{w}(x_{h}(t)) - 
\varphi_{w}(x_{e}(t))], 
\end{equation} 
\begin{equation} 
\label{e:ramoi} 
I(t) = Q_{0} \cdot [\vec{E}_{w}(x_{h}(t)) \cdot 
\vec{v}_{h}(t) - \vec{E}_{w}(x_{e}(t)) \cdot 
\vec{v}_{e}(t)], 
\end{equation} 
where $Q_{0}$ is the electric charge carried by electrons or holes, 
$x_{e/h}(t)$ and $\vec{v}_{e/h}(t)$ are the position and 
velocity vectors of electrons/holes as a function of time, and 
$\varphi_{w}$ and $\vec{E}_{w}$ are the weighting potentials 
and weighting fields.

Figure~\ref{f:ps}a shows the weighting potential of a segment, B,  
with the 
indication of a photon interaction. 
The drift trajectories of holes and 
electrons are also indicated. Figures~\ref{f:ps}\,(b) and (c) show 
the charges and currents induced on the electrodes of the segment
with the interaction and its neighbors.
The 
pulses induced in the neighboring segments are called mirror 
pulses. The amplitude of the mirror pulse induced in segment~A is  
larger than the one in segment~C since the trajectory 
is closer to segment~A than~C.

The charge pulses in segment~B and the core show clear kinks at the time when
the inwards drifting electrons reach the core electrode, i.e. the inner
mantle. At that time the current pulses show maxima. The charge mirror-pulses
do not show clear features at this time; the current mirror-pulses
show more distinct features.

The position in $r$ of an energy deposit is clearly reflected in the
resulting pulses in the segment and core electrodes. This is demonstrated
in Figs.~\ref{f:psrad1} and~\ref{f:psrad2}.

\section{Influence of electronics} 
\label{s:dbn} 

The most widely used amplification scheme is based on
charge sensitive devices. The bandwidth is limited by the
amplifiers and by the transmission cables. 
In order to account for the limited bandwidth and the decay time 
of the amplifier, 
the simulated pulses can be convoluted with 
a transfer function representing the readout electronics.  
In addition, Gaussian noise can be added 
to represent the noise of the system.
 
\begin{figure}[htpb] 
\centering 
\includegraphics[width=0.9\linewidth]{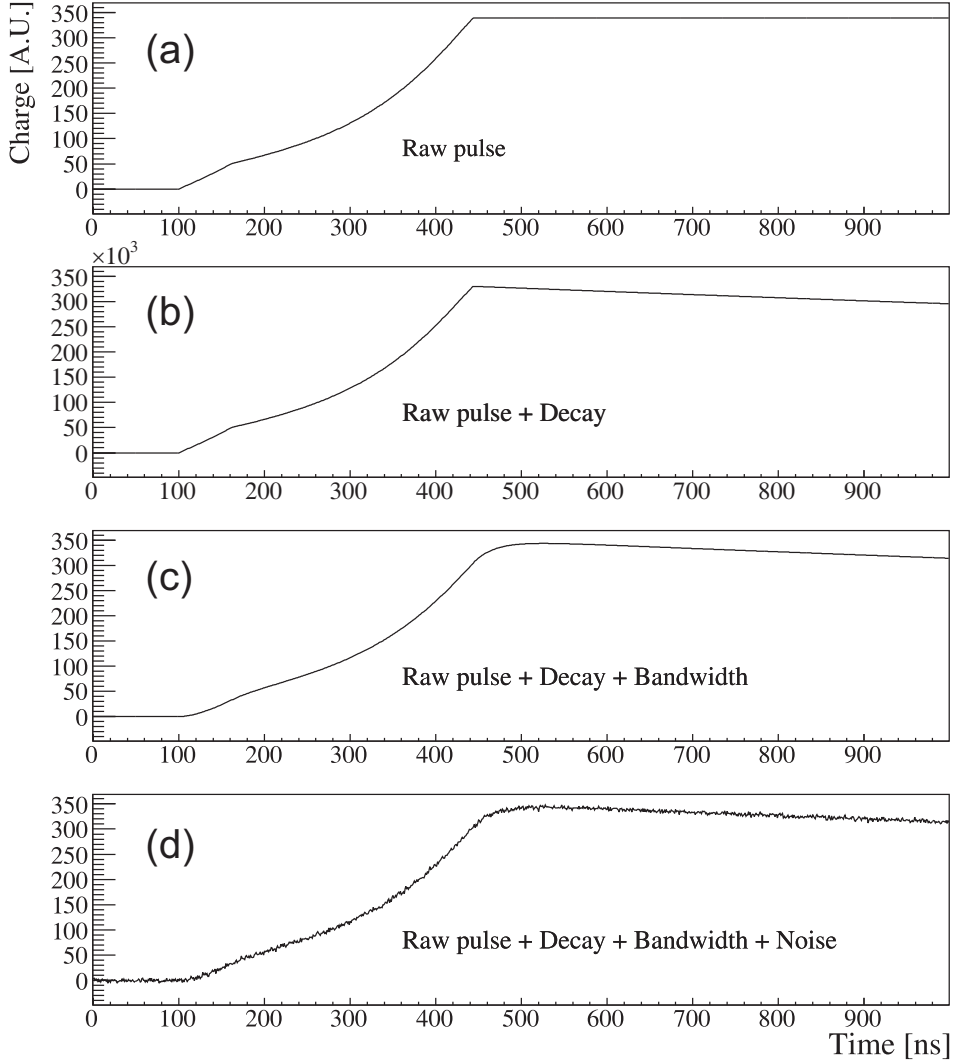} 
\caption{Pulses at different stages of the simulation: (a)Raw pulse, 
(b) after folding in in the decay time of the signal,(c) after convolution 
with the bandwidth and (d) after adding noise.} 
\label{f:elec} 
\end{figure}

 
Figure~\ref{f:elec} demonstrates the modification of a simulated pulse 
after folding in a decay time of $5\,\mu$s and a 
cut-off in bandwidth was 10~MHz.
The noise level was set to
5\,\% of the pulse amplitude. These noise and band-width values 
are chosen to clearly demonstrates the effects.

The decay-time of the preamplifier was chosen significantly larger than
the length of the pulse. Therefore, mainly the plateau is affected.
The cut-off in bandwidth smoothes the features of the pulse.
The kink clearly visible at $t$\,=\,150\,ns in Figure~\ref{f:elec}(b)
is washed out significantly. The noise applied for Figure~\ref{f:elec}(d)
makes the kink almost invisible.

\section{Uncertainties of the simulation}
\label{s:unc}

\begin{figure}[htpb] 
\centering 
\includegraphics[width=0.9\linewidth]{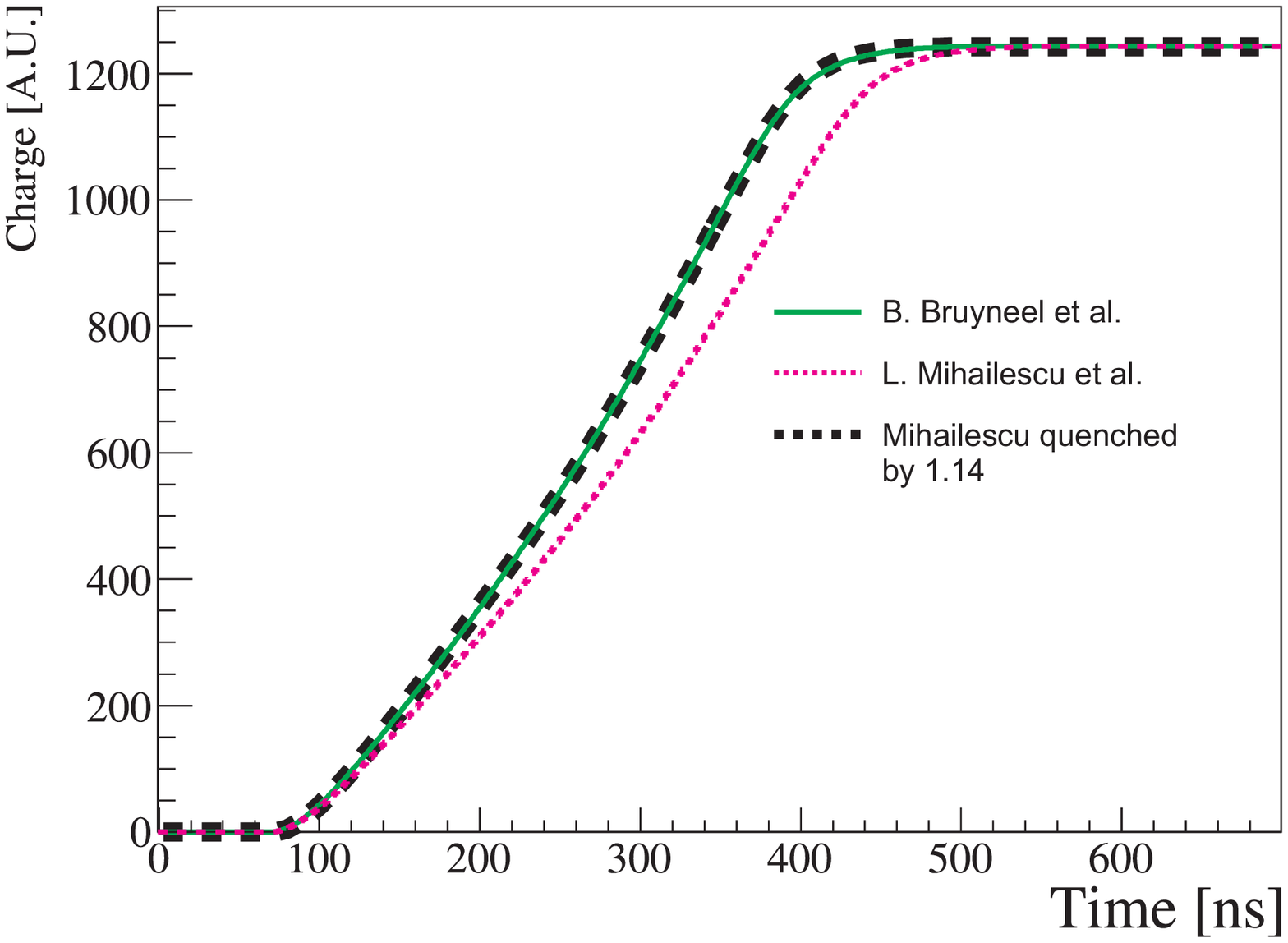} 
\caption{Simulated core pulses for two different sets
of input parameters for the electron mobility. The result of a time scaling
of the longer pulse is also shown.}
\label{f:cmob} 
\end{figure}
 
\begin{figure}[htpb] 
\centering 
\includegraphics[width=0.9\linewidth]{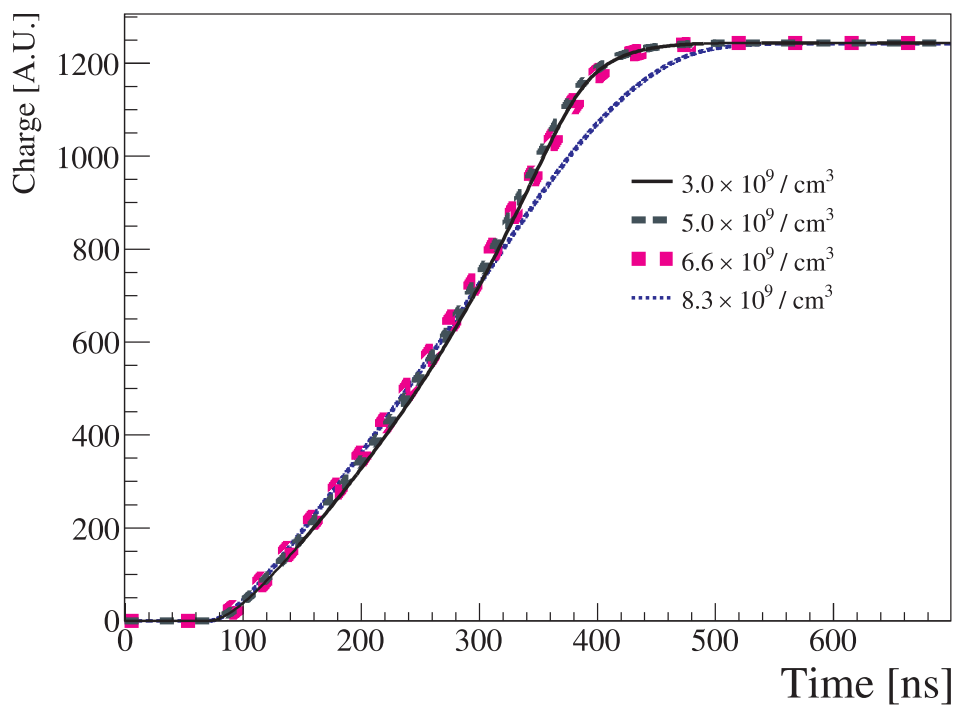} 
\caption{Simulated core pulses for different impurities, $\rho_{imp}$.}
\label{f:cimp} 
\end{figure}

The output of the simulation depends critically on the input parameters
used. 
A fundamental problem is the input concerning the
mobilities. Figure~\ref{f:cmob} demonstrates how the pulses 
of events starting at the outer mantle and observed in the core
are stretched if a different set~\cite{miha} of
input parameters for the electron mobility is chosen. 
At $V_{bias}\,=$\,3000\,V
and $\rho_{imp}\,=\,0.66\,\cdot\,10^{10}/$cm$^3$, the pulse
is 14\,\% longer than with the default input parameters~\cite{bart}.
However, a simple time scaling makes the pulses indistinguishable.

The length of a pulse is also strongly influenced by $\rho_{imp}$.
This is demonstrated in Fig.~\ref{f:cimp}. 
At $V_{bias}\,=$\,3000\,V a change in
$\rho_{imp}$ from $\,0.66\,\cdot\,10^{10}$ to
$\,0.83\,\cdot\,10^{10}/$cm$^3$
has an effect on the core pulse of similar size
than changing the  mobility.
In this case, the shape of the pulse is also modified;
a simple time scaling does not make the pulses inddistinguishable.
However, the effect is quite subtle and will be difficult
to observe in real data.
Unfortunately, the value of  $\rho_{imp}$ is generally not known
to very high precision. In addition, it often varies 
up to a factor of three along $z$. 
Therefore, it is unavoidable
that pulses simulated for a particular detector or even detector section
have to be scaled in time before  
more detailed comparisons to data can be made. 
Assuming a certain mobility, the length of the measured pulses 
can be used to derive the impurity density locally.

\section{Comparison to data} 
\label{s:psv} 

The focus of the comparisons presented here is the
drift of electrons. Therefore pulses as seen in the core
electrode were used.
Simulated pulses were compared to measured pulses produced by
a well understood true-coaxial $n$-type detector~\cite{si}
irradiated with a collimated Europium source.
The 122\,keV line was used to study energy deposits close to the
detector outer mantle. In this case, the holes are quickly absorbed
by the electrodes on this outer mantle and mainly the drift of the
electrons is relevant for the shape of the resulting pulses.

Two methods were used. Simulated pulses were either fitted to
individual measured pulses or to averages of sets of pulses
obtained in a given configuration. The latter was done to
reduce the effects of noise.  

\subsection{Test environment}

\begin{figure}[htbp] 
\centering 
\includegraphics[width=0.8\linewidth]{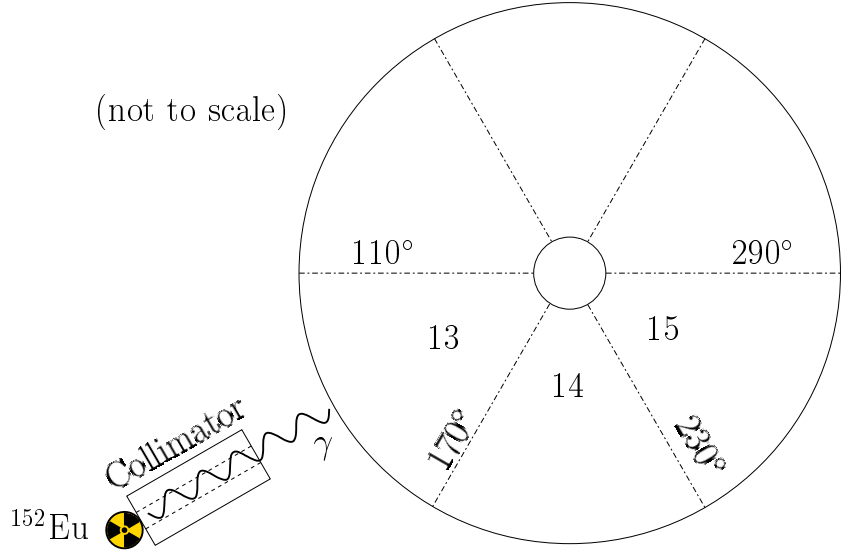} 
\caption{Schematic of the setup used to scan three segments of an
$(6\phi,3z)$ segmented detector at $z=0$}
\label{f:siscan} 
\end{figure}

Figure~\ref{f:siscan} depicts the measurement setup.
Three segments, numbered 13 to 15, of the middle layer 
of the detector were scanned in $\phi$ at $z=0$.  
A collimated 75\,kBq~$^{152}$Eu source inside a copper collimator
was used.
The collimator pointed to the center of the detector and
the spot had a 1\,$\sigma$ diameter of about 5\,mm on the 
outer mantle of the detector. 
Details of the setup can be found elsewhere~\cite{si}. 
The operational voltage was 3\,kV. 
The step size of the scan was 
5$^{\circ}$ in segment 14 and 10$^{\circ}$ in segments 13 and 15. The 
uncertainty in $\phi$ was approximately $\Delta \phi = 2.5^{\circ}$. 
In total, 25~steps were made to cover 180$^{\circ}$ in 
azimuth. The pulses of the core and all segments were recorded.  
Events from the 122~keV $\gamma$-line were 
selected with a cut on the core energy of 2\,$\sigma$~of the energy
resolution, i.e. about $\pm$~4keV.

Simulations were done with the energy deposits  located exactly
on the outer mantle or  distributed according to the
interactions of the 122\,keV photons from the $^{152}$Eu source. 
A constant  $\rho_{imp}\,=\,1.05\,\cdot\,10^{10}/$cm$^3$ was implemented.
This is the average of the values given by the manufacturer
for the top and the bottom of the detector which differ by a factor
of two.
The electric field was calculated for a bias voltage of  $V_{bias}=3\,$kV.
The measured orientation
of the crystal axes was used in the simulation; 
the $\langle 110 \rangle$ axis was set to $290^\circ$, almost aligned
with one boundary of segment~15. 

An amplifier decay-time of 50~$\mu$s and a cut-off in 
bandwidth of 37.5\,MHz were implemented according to the specifications 
of the electronics system. Electronic noise was not added to the 
simulated pulses to simplify direct comparisons with individual 
measured pulses.

\subsection{Fits to individual pulses} 

Simulated core pulses, $C_{sim}(t)$, were fitted to measured
pulses, $C_{meas}$, using three parameters only: the amplitude,
$A$, the time offset, $T_0$, and the time scaling-factor    
$T_{scale}$:

\begin{equation} 
\label{e:tsc} 
C_{meas}(t) = A \times C_{sim}(t / T_{scale}  + T_0) \,\,\,. 
\end{equation}

\begin{figure}[htbp] 
\centering 
\includegraphics[width=0.9\linewidth]{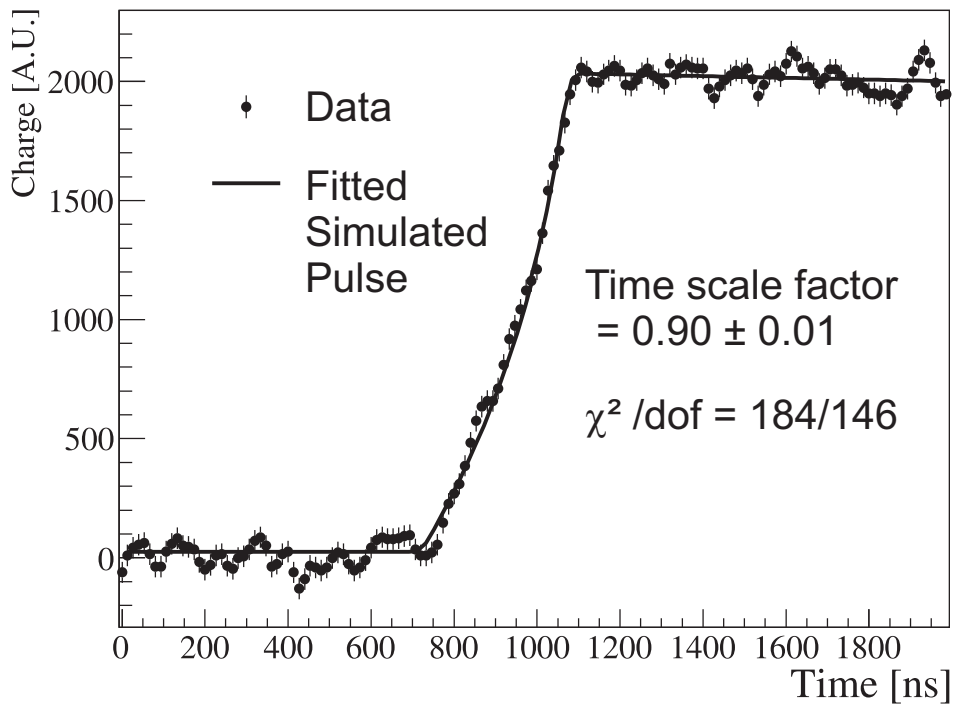} 
\caption{Fit of a simulated to a measured pulse at $\phi=200^{\circ}$. 
The dots represent 
the data with error bars indicating the noise level.} 
\label{f:s2d} 
\end{figure}

Figure~\ref{f:s2d} shows a randomly selected core pulse at $\phi=200^{\circ}$
and the fit of the pulse generated for this angle.  
This position falls 
between the $\langle 110 \rangle$ and the $\langle 100 \rangle$ axes. 
The actual values of the parameters~$A$ and  $T_0$ have no relevance; 
they are not simulated in real units.
The important parameter is $T_{scale}$ which has a value of 
0.90\,$\pm$\,0.01. The $\chi^{2}$/dof\,=\,184/146\,=\,1.3 
for this fit is quite good.

\begin{figure}[htbp] 
\centering 
\includegraphics[width=0.9\linewidth]{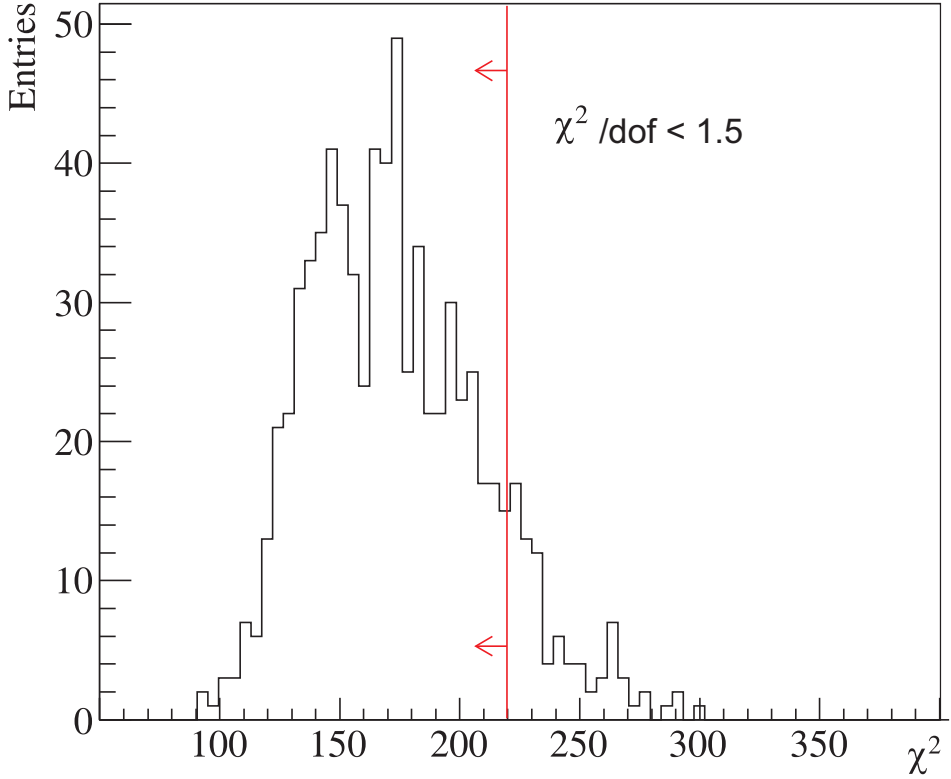} 
\caption{$\chi^{2}$ distribution from the fits of the simulated to 
measured pulses taken at $\phi=200^{\circ}$. Only events with 
$\chi^{2}$/dof$\,<\,1.5$  are analysed further. 
The cut is indicated by a vertical line.} 
\label{f:x2} 
\end{figure}

Fig. \ref{f:x2} shows the distribution of $\chi^{2}$/dof values 
of all fits of pulses at $\phi=200^{\circ}$. The average 
$\chi^{2}$/dof\,=\,174/146\,=\,1.2.
In general, the quality of the fits is very good. This indicates that
the simulation describes the general shape of the pulses very well. 
To determine an overall $T_{scale}$, a clean subset  
of measured pulses was selected by:
\begin{itemize} 
\item requiring $\chi^{2} < 220$, i.e. $\chi^{2}$/dof\,$<\,1.5$,
      to eliminate background events which have
      probably more than one interaction in the segment; 
\item rejecting events in which an intermittent DAQ problem
      might have affected the recording of the pulse. 
\end{itemize} 
 
 
\begin{figure}[htbp] 
\centering 
\includegraphics[width=0.9\linewidth]{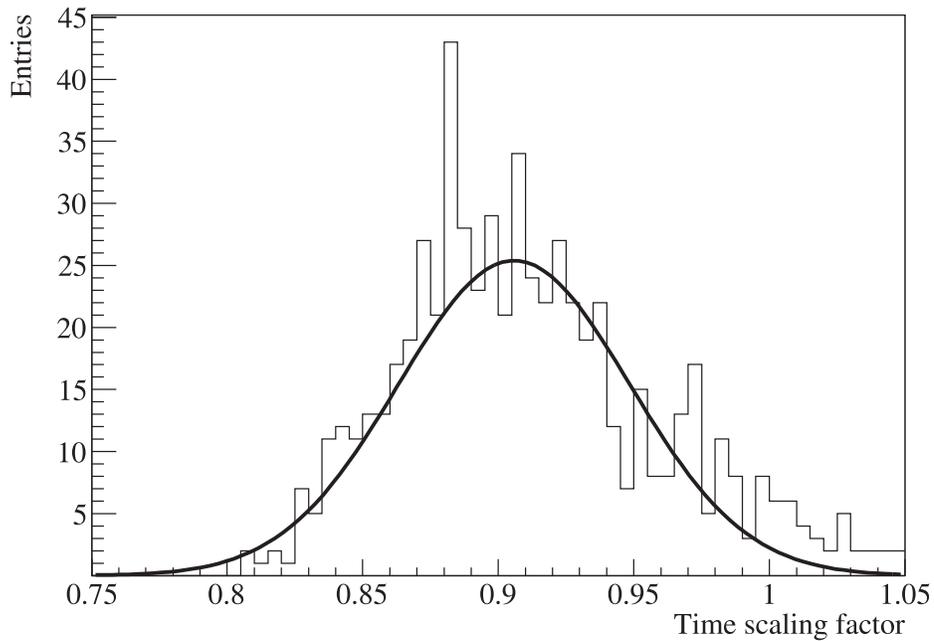} 
\caption{Distribution of time scaling-factors  
from the fits of the simulated to te
measured pulses taken at $\phi=200^{\circ}$.
The result of a Gaussian fit is shown as a solid line.} 
\label{f:ts200} 
\end{figure}

\begin{figure}[htpb] 
\centering 
\includegraphics[width=0.9\linewidth]{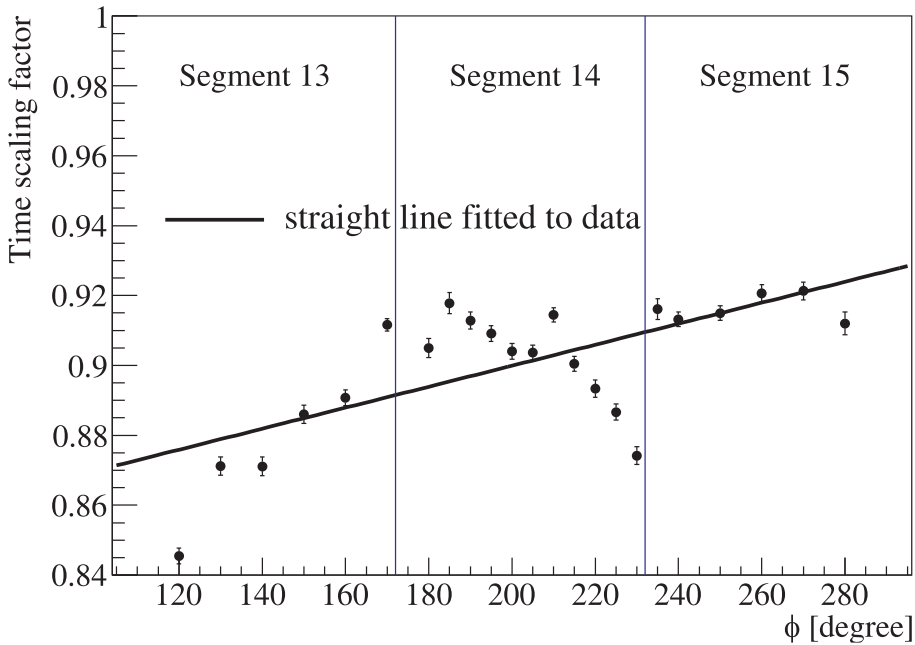} 
\caption{Mean time scaling-factors, points, as extracted for
         each $\phi$. The uncertainties were taken from the Gaussian fits.
         The result of a straight 
         line fit to the points is given as a solid line.} 
 
\label{f:tsl} 
\end{figure}

For each pulse, a value of $T_{scale}$ was determined.
Figure~\ref{f:ts200} shows the distribution of the 
time scaling-factors of the pulses at $\phi=200^{\circ}$.
The mean time scaling-factor of $0.91\,\pm\,0.01$ 
was extracted using  a Gaussian fit.
This was done for all positions in $\phi$.
Figure~\ref{f:tsl} shows the $\phi$ dependence of the 
mean time scaling-factors. 
The dotted vertical lines indicate the segment boundaries.  
A value of $T_{scale}=1$, independent of $\phi$
would be expected for a perfect simulation.
However, considering the uncertainties in the input parameters,
the level of agreement is quite astonishing.
The time scaling-factors are systematically about 10\,\% low.
This means that the simulated pulses are too short, indicating that
the velocities of the electrons are overestimated.
As discussed in section~\ref{s:unc}, this could be due to
too high mobilities assumed or due to a slightly underestimated
$\rho_{imp}$.
It is impossible to distinguish the two possibilities at this point.
 
The $\phi$~dependence was fitted with a straight line. 
The absence of an oscillation pattern with a period of 90$^{\circ}$ 
shows that the longitudinal anisotropies introduced in the model
describe reality in a satisfactory manner.
The obvious dependence of $T_{scale}$ on $\phi$ is difficult to explain.
In germanium crystals, $\rho_{imp}$ often depends on $r$. If 
the central $z$-axis of the original crystal is not identical with the
$z$-axis of the detector, 
a $\phi$ dependence of the observed average  $\rho_{imp}$
can result. A further investigation whether this effect can explain
the about 5\,\% effect is under way.

\subsection{Fits to averaged pulses}

\begin{figure}[htbp]
\centering
\includegraphics[width=0.9\linewidth]{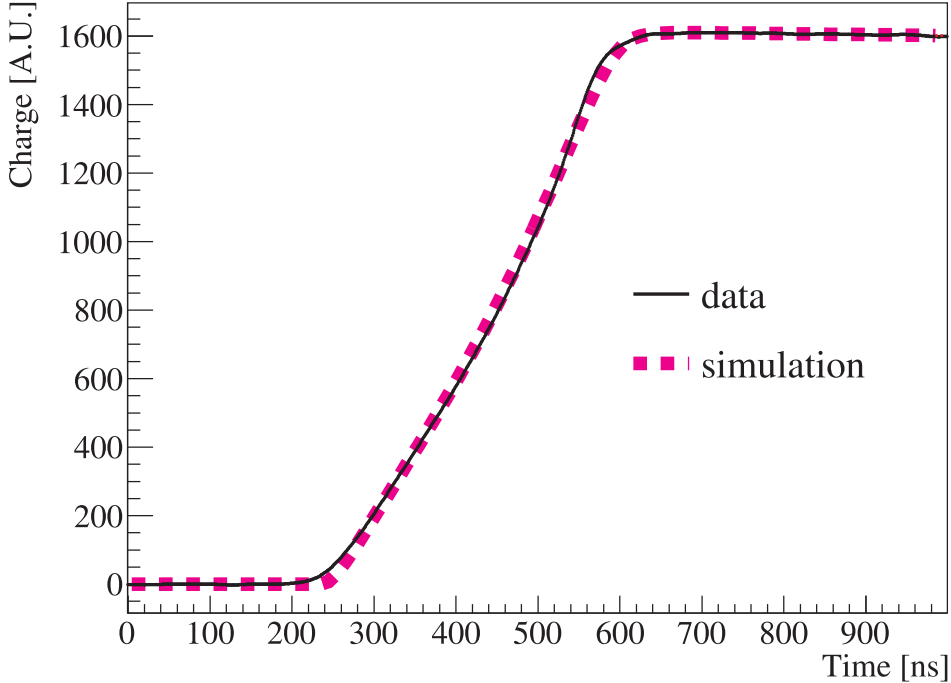}
\caption{Fit of the averaged simulated to the averaged measured
pulse at $\phi=200^{\circ}$.
The statistical uncertainties in the
averaged measured pulse are too small to be seen.
}
\label{f:ave}
\end{figure}

It would be interesting to carefully study the detailed shapes 
of individual pulses to distinguish between different sources
of deviation between simulation and data.
However, Fig.~\ref{f:s2d} shows that the noise level in this measurement
does not allow this.
In order to remove the effects of noise,
the selected pulses at each scan position were averaged, even though
the interaction points were spatially distributed according to the beam spot
and the penetration depth of 122\,keV photons.

The simulation was adjusted accordingly and
the simulated pulses were also averaged. 
Averaged simulated pulses were fitted to averaged measured pulses.
The result for $\phi\,=\,200^{\circ}$ is shown in Fig.~\ref{f:ave}. 
The fit yields $T_{scale}\,=\,0.90\,\pm\,0.01$.
This is compatible to the mean $T_{scale}\,=\,0.91\pm\,0.01$ observed before.

The $\chi^{2}$/dof of the fit is $\approx$\,8.
This is quite large. In Fig.~\ref{f:ave}, the deviations are also clearly 
visible. The simulated pulse has sharper edges 
most likely due to the overly optimistic cut
of 37.5\,MHz
in the bandwidth; Figure~\ref{f:elec} demonstrates how pulses get smoothed
out by a cut in bandwidth of 10\,MHz.
At the very end of the pulse,
the simulation shows an extra flattening.
This would indicate that $\rho_{imp}$ is slightly
smaller than assumed.
Nevertheless, the overall shape of the pulse is reproduced rather 
well and it will be tested soon, how well simulated pulses work
as training sets for neural network based pulse shape analyses.

\section{Summary and Outlook}
\label{s:sum}

A fully functional pulse shape simulation package 
for HPGe detectors was developed.
It can easily be adapted to
any common HPGe detector geometry.
The results of the simulation are critically dependent on the
input parameters.
While some parameters are generic to germanium, the detailed properties
of a detector like impurity concentration and axes orientation
are also extremely important. Simulations have to be adjusted
to each detector.

The longitudinal anisotropy of the elecron drift is correctly
described by the simulation.
Therefore the comparison of simulated to measured pulses can yield
information about intrinsic detector properties like local variations
of the impurity concentration.

The quality of the simulated pulses will be further studied and
the simulation will be used as input to pulse shape analyses.

\section{Acknowledgements}

We would like to thank the members of the Monte Carlo group of the Majorana
collaboration for their kind help in the analysis and cooperation in
the programming.



\end{document}